\documentclass[aps,longbibliography,superscriptaddress,twocolumn,nofootinbib]{revtex4-1}
\usepackage{graphicx}
\usepackage{stmaryrd}
\usepackage{amsmath, amsfonts,amssymb}
\usepackage{mathrsfs}
\usepackage{braket}
\usepackage{color}
\usepackage{xspace}
\usepackage{amscd}
\usepackage{comment}
\usepackage{bm}
\definecolor{lightblue}{rgb}{0.13, 0.26, 0.99}

\usepackage[
colorlinks=true,
urlcolor=blue,
citecolor=blue,
linkcolor=blue,
hyperfootnotes=false]{hyperref}

\allowdisplaybreaks

\begin{document}

\title{Anomalies of kagome antiferromagnets on magnetization plateaus}
\author{Shunsuke C. Furuya}
\affiliation{Department of Physics, Ibaraki University, Mito, Ibaraki 310-8512, Japan}
\affiliation{Condensed Matter Theory Laboratory, RIKEN, Wako, Saitama 351-0198, Japan}
\author{Yusuke Horinouchi}
\affiliation{RIKEN Center for Emergent Matter Science (CEMS), Wako, Saitama 351-0198, Japan}
\author{Tsutomu Momoi}
\affiliation{Condensed Matter Theory Laboratory, RIKEN, Wako, Saitama 351-0198, Japan}
\affiliation{RIKEN Center for Emergent Matter Science (CEMS), Wako, Saitama 351-0198, Japan}

\begin{abstract}
    We discuss the ground-state degeneracy of spin-$1/2$ kagome-lattice quantum antiferromagnets on magnetization plateaus by employing two complementary methods: the adiabatic flux insertion in closed boundary conditions and a 't Hooft anomaly argument on inherent symmetries in a quasi-one-dimensional limit.
    The flux insertion with a tilted boundary condition restricts the lower bound of the ground-state degeneracy on $1/9$, $1/3$, $5/9$, and $7/9$ magnetization plateaus under the $\mathrm{U(1)}$ spin-rotation and the translation symmetries: $3$, $1$, $3$, and $3$, respectively.
    This result motivates us further to develop an anomaly interpretation of the $1/3$ plateau.
    Taking advantage of the insensitivity of anomalies to spatial anisotropies, we examine the existence of the unique gapped ground state on the $1/3$ plateau from a quasi-one-dimensional viewpoint.
    In the quasi-one-dimensional limit, kagome antiferromagnets are reduced to weakly coupled three-leg spin tubes.
    Here, we point out the following anomaly description of the $1/3$ plateau. While a simple $S=1/2$ three-leg spin tube cannot have the unique gapped ground state on the $1/3$ plateau because of an anomaly between a $\mathbb Z_3\times \mathbb Z_3$ symmetry and the translation symmetry at the $1/3$ filling, the kagome antiferromagnet breaks explicitly one of the $\mathbb Z_3$ symmetries related to a $\mathbb Z_3$ cyclic transformation of spins in the unit cell.
    Hence the kagome antiferromagnet can have the unique gapped ground state on the $1/3$ plateau.
\end{abstract}

\date{\today}
\maketitle

\section{Introduction}\label{sec:intro}

The last decade has witnessed the bloom of topological classification of gapped quantum phases~\cite{schnyder_top, chen_class_1d, chen_spt_cohom}.
Topology, in a broad sense, has made possible the universal classification of gapped topological phases outside the reach of the order-parameter paradigm.
By contrast, much less is known about the classification of gapless quantum phases protected by symmetries~\cite{furuya_wzw, yao_lsm_anomaly}.
One milestone in the classification of gapless quantum phases is the Lieb-Schultz-Mattis (LSM) theorem originally proven for spin-$1/2$ XXZ chains stating that they cannot have a unique gapped ground state~\cite{lsm}.
The LSM theorem is related to the $\mathrm{U(1)}$ flux insertion to quantum many-body systems, where the filling of particles plays an essential role.
A fractional filling is necessary, not sufficient, though, for excluding the possibility of the unique gapped ground state in the periodic boundary condition.
On the basis of the filling argument, the LSM theorem was extended to quantum many-body systems in two or higher dimensions~\cite{oshikawa_lsm, hastings_lsm}.

In quantum spin systems, the LSM theorem is immediately related to magnetization plateaus, as exemplified by the Oshikawa-Yamanaka-Affleck (OYA) condition~\cite{oya_plateau}.
The OYA condition successfully explains the possible emergence of plateaus and their ground-state degeneracy in (quasi-)one-dimensional quantum spin systems with the filling argument in analogy with that of the LSM theorem.
Later the OYA condition was extended to two or higher dimensional quantum spin systems by Oshikawa himself~\cite{oshikawa_lsm}.
Interestingly, the adiabatic flux insertion argument in Ref.~\cite{oshikawa_lsm} enables us to deal with the OYA condition on equal footing with the LSM theorem.
However, the flux-insertion argument in Ref.~\cite{oshikawa_lsm} leads to a system-size dependent ground-state degeneracy, which makes the thermodynamic limit ambiguous.
The system-size dependence originates from the periodic boundary condition in one direction, where the $d$-dimensional system can be regarded as a product $S^1\times M_{d-1}$ of a  ring $S^1$ and a $(d-1)$-dimensional ``cross-section'' $M_{d-1}$.
When the argument of Ref.~\cite{oshikawa_lsm} is applied to quantum spin systems,
the number of spins on $M_{d-1}$ must not be any integral multiple of the filling fraction to exclude the possibility of the unique gapped ground state, leading to the inconvenient system-size dependence of the ground-state degeneracy.
Still, the flux insertion argument is advantageous even today for its intuitive picture and its direct connection to 't Hooft anomalies~\cite{cho_anomaly_lsm,yao_lsm_boundary, furuya_checkerboard}, which would possibly be a counterpart of the topology in the classification of gapless phases.

The importance of anomalies in condensed matter physics has been well recognized, for example, in the context of symmetry-protected surface states of topological phases~\cite{ryu_mod_inv, sule_spt_orb}.
It was also pointed out that the LSM-type argument of $d$-dimensional bulk phases is related to surface anomalies of ($d$+1)-dimensional weak SPT phases~\cite{meng_set_surface, Jian_LSM_anomaly, thorngren_crystalline}.
However, such an anomaly description of magnetization plateaus in two or higher dimensions is yet to be established.
Unambiguous flux insertion argument of magnetization plateaus will offer the first step toward their anomaly description.

Recently, the authors~\cite{furuya_checkerboard} derived the OYA condition in frustrated quantum magnets on the checkerboard lattice by adapting the $\mathrm{U(1)}$ flux insertion argument avoiding the problem in the periodic boundary condition.
The key idea in this approach is to use a closed boundary condition accompanied by a spatial twist that preserves the checkerboard symmetries instead of the simple periodic one.
This argument is based on a simple assumption that we can choose arbitrary boundary conditions if they keep the symmetries in question.
Reference~\cite{furuya_checkerboard} provides us with a viewpoint that an appropriate symmetric boundary condition gives shape to a relation between the magnetization plateau and the anomaly.
However, the spatially twisted boundary condition stands on characteristics of the checkerboard and is thus inapplicable to a broad class of geometrically frustrated quantum spin systems with triangle-based lattices.
Most importantly, it is inapplicable to kagome-lattice quantum antiferromagnets, which are famous for their fertility of magnetization plateaus~\cite{hida_kagome_mag, schulenburg_kagome_mag, nishimoto_kagome_dmrg, Capponi_kagome_plateau, nakano_1/3, nakano_spin-s, picot_kagome}.

In this paper, we revisit magnetization plateaus of $S=1/2$ quantum spin systems on the kagome lattice and investigate them from the viewpoint of the OYA condition in a symmetric closed boundary condition such as a tilted boundary condition~\cite{yao_lsm_boundary}.
Some closed boundary conditions, including the tilted one, enable us to regard kagome quantum antiferromagnets as one-dimensional quantum antiferromagnets with long-range interactions.
This viewpoint gives us some insight into the $1/3$ magnetization plateau.
If we consider a simple three-leg spin tube with short-range interactions~\cite{sakai_tube, okunishi_tube_chiral, plat_tube, yonaga_tube_twisted}, 
we find a 't Hooft anomaly in a (1+1)-dimensional quantum field theory as an effective description of the $S=1/2$ three-leg spin tube on the $1/3$ magnetization plateau.
The 't Hooft anomaly of this field theory excludes the possibility of the unique gapped ground state on the $1/3$ plateau of the spin tube under a certain on-site symmetry and the translation symmetry.
However, when considered as a three-leg spin tube the kagome antiferromagnet has a long-range interaction, which hinders the direct inheritance of the anomaly of the simple three-leg spin tube.

To discuss how the anomaly of the three-leg spin tube is broken in the kagome antiferromagnet on the $1/3$ plateau,
we take another approach from a quasi-one-dimensional limit.
We can see the absence of the anomaly more explicitly by introducing a spatial anisotropy to the Hamiltonian because the anisotropy preserves the required symmetries.
In the quasi-one-dimensional limit, the kagome antiferromagnet turns into a weakly coupled three-leg spin tubes with short-range interactions.
Here, we can find that the kagome lattice's symmetry explicitly breaks one of these symmetries involved with the anomaly unless the Hamiltonian is fine-tuned.
The quasi-one-dimensional viewpoint tells us that the unique gapped ground state on the $1/3$ plateau is permitted by the kagome geometry thanks to the resolution of the anomaly.

The paper is organized as follows.
We define a symmetric closed boundary condition called the tilted boundary condition~\cite{yao_lsm_boundary} in Sec.~\ref{sec:tilt} and argue the flux insertion with this boundary condition.
In Sec.~\ref{sec:1d}, we develop an anomaly argument on magnetization plateaus of kagome antiferromagnets as coupled spin tubes.
After these sections, we summarize this paper in Sec.~\ref{sec:summary}.

\section{Tilted boundary condition}\label{sec:tilt}

\begin{figure}
    \centering
    \includegraphics[viewport = 0 0 1300 800, width=\linewidth]{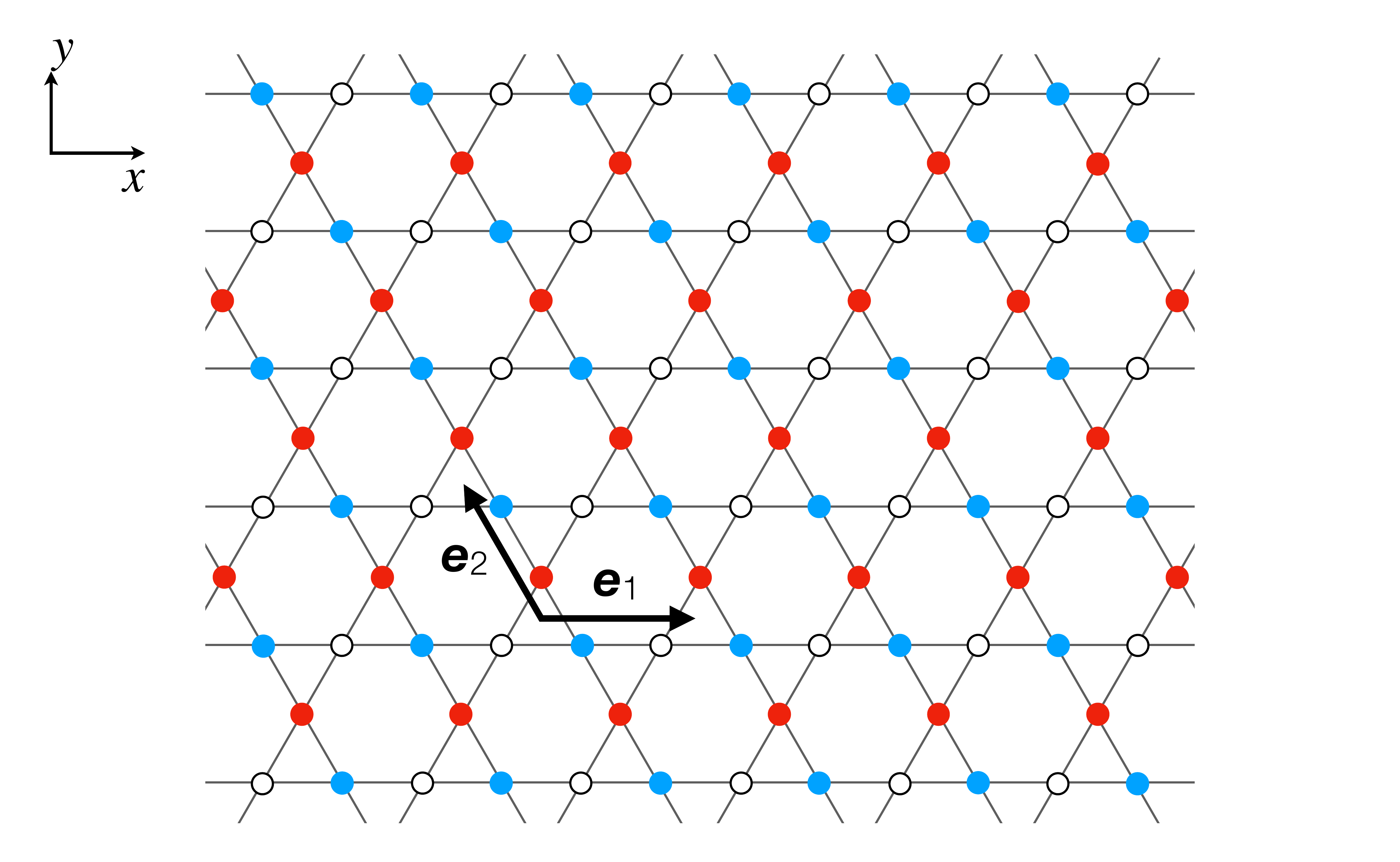}
    \caption{The kagome lattice. 
    The location of the unit cell is specified by a two-dimensional vector $\bm R = n_1 \bm e_1 + n_2 \bm e_2$ with $n_1, n_2\in \mathbb Z$ and two-dimensional unit vectors $\bm e_1$ and $\bm e_2$.}
    \label{fig:kagome}
\end{figure}

\subsection{Assumption}

Let us begin by clarifying an assumption about the effect of boundary conditions that we rely on in this paper.
The assumption is that if a system has the unique gapped ground state in the periodic boundary condition under certain symmetries, it also does in any other closed boundary conditions that respect the symmetries.
This assumption is natural since the closed boundary condition is just an artificial condition of theories to minimize the boundary effect.
The bulk properties should be independent of a specific choice of symmetric boundary conditions.

Taking the contraposition of the assumption, we can state that if the unique gapped ground state is forbidden in one symmetric closed boundary condition,  it is also forbidden in the periodic boundary condition.
This statement motivates us to search for an appropriate symmetric closed boundary condition that clarifies a universal constraint forbidding the unique gapped ground state.
Among such symmetric closed boundary conditions is the tilted boundary condition~\cite{yao_lsm_boundary} that we consider in this section.

\subsection{Symmetries}

Before defining the tilted boundary condition, we define the model and its symmetries that we use in this paper.
We discuss spin-$1/2$ kagome antiferromagnets with translation symmetries and the $\mathrm{U(1)}$ spin-rotation symmetry.
We assume translation symmetries of a unit cell by one unit in the $\bm e_1$ and the $\bm e_2$ directions of Fig.~\ref{fig:kagome}, where the smallest upward triangle is considered as the unit cell.
All the results derived in this section also hold for spin-$S$ models with $S> 1/2$.

Let us specify the location of the unit cell by a two-dimensional vector $\bm R= n_1 \bm e_1 +n_2 \bm e_2$ with $\bm e_1 = (1,0)$, $\bm e_2 = (-\frac 12, \frac{\sqrt 3}{2})$ (Fig.~\ref{fig:kagome}), and $n_1, n_2 \in \mathbb Z$.
We employed a unit of the lattice spacing $a=1$.
Accordingly, the spin operator can be denoted as $\bm S_{\mu}(n_1,n_2)$.
In the figures, the indices $\mu=1,2$, and $3$ are distinguished visually by red, white, and blue circles, respectively.
Then the Hamiltonian of the spin-$S$ Heisenberg antiferromagnet is represented as
\begin{align}
    \mathcal H
    &= J \sum_{n_1,n_2} [\bm S_1(n_1,n_2) \cdot \bm S_2(n_1,n_2) 
    \notag \\
    & + \bm S_2(n_1,n_2) \cdot \bm S_3(n_1,n_2)  + \bm S_3(n_1,n_2) \cdot \bm S_1(n_1,n_2) ]
    \notag \\
    & + J \sum_{n_1,n_2} [ \bm S_3(n_1,n_2) \cdot \bm S_2(n_1+1,n_2)
    \notag \\
    &+ \bm S_1(n_1,n_2) \cdot \{ \bm S_3(n_1, n_2+1) + \bm S_2(n_1+1, n_2+1) \} ]
    \notag \\
    &- h \sum_{n_1,n_2,\mu} S_\mu^z(n_1,n_2),
    \label{H_kagome_HAFM}
\end{align}
with $J>0$ and $h\ge 0$.
This uniform kagome Heisenberg antiferromagnet is merely a specific example that satisfies the $\mathrm{U(1)}$ and the $T_1$ and $T_2$ translation symmetries, where $T_n$ represents the translation by the one unit in the $\bm e_n$ direction:
\begin{align}
    T_1 \bm S_\mu(n_1,n_2) T_1^{-1} &= \bm S_\mu(n_1+1,n_2),
    \label{T1} \\
    T_2 \bm S_\mu (n_1,n_2) T_2^{-1} &= \bm S_\mu (n_1,n_2+1).
    \label{T2}
\end{align}
We can add any symmetric interactions to the Hamiltonian whenever we want.

To make the translation symmetries well-defined, we need to specify the boundary condition.
The Hamiltonian \eqref{H_kagome_HAFM} possesses the $\mathrm U(1)$ spin-rotation symmetry and translation symmetries if the periodic boundary condition is imposed on the $x$ and the $y$ directions (Fig.~\ref{fig:kagome}).
When we adopt the flux insertion argument~\cite{oshikawa_lsm} to this system with the periodic boundary condition, we face the previously mentioned problem of the ambiguous thermodynamic limit~\cite{Pal_kagome}.
In the case of the $d$-dimensional hyper cubic lattice, this problem can be resolved in Ref.~\cite{yao_lsm_boundary} by another closed boundary condition that respects the symmetries, the tilted boundary condition.

\subsection{Tilted boundary condition}

\begin{figure}[t!]
    \centering
    \includegraphics[viewport=0 0 1300 800, width=\linewidth]{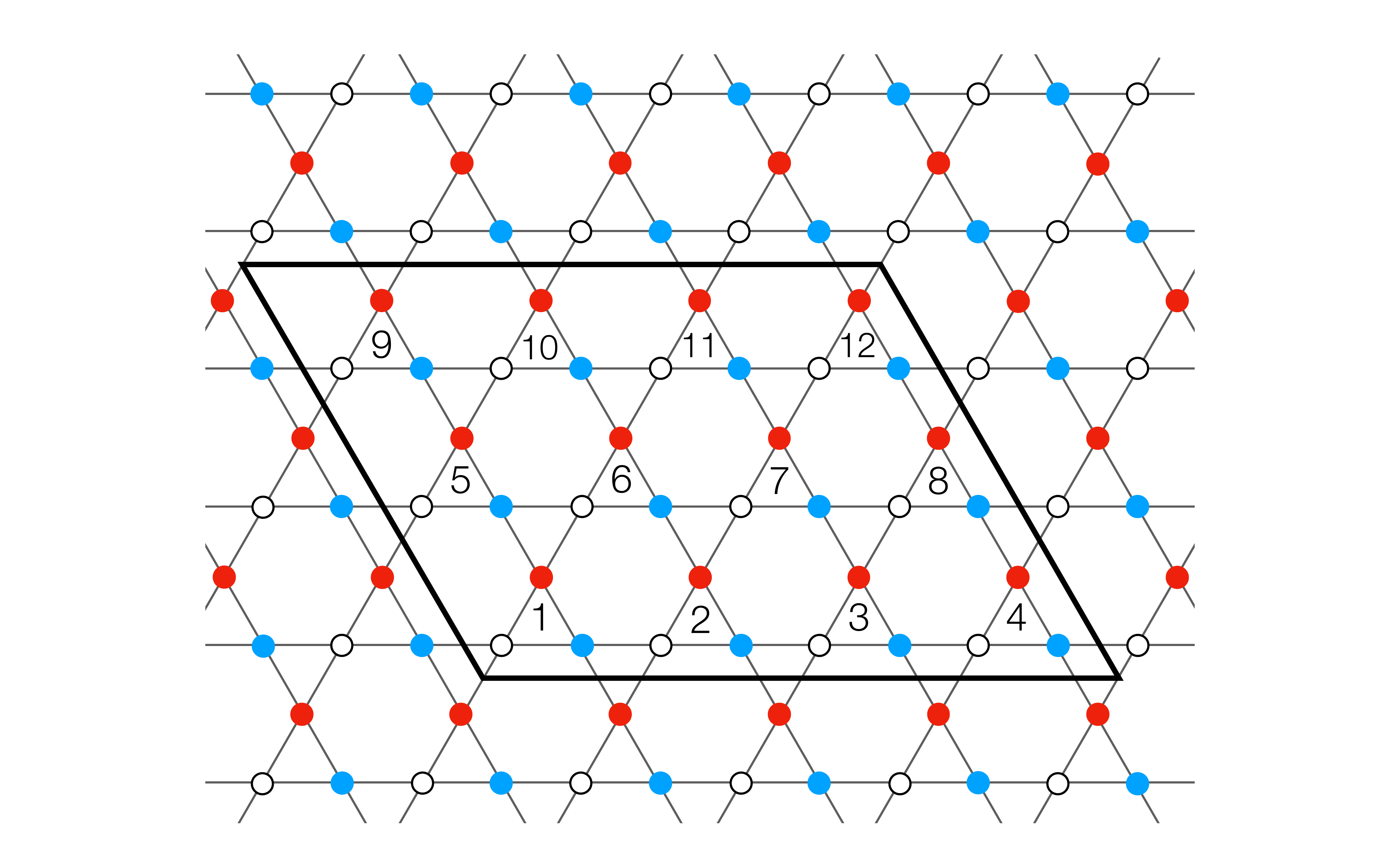}
    \caption{A rhombus finite-size cluster of the kagome lattice with $36$ sites in the tilted boundary condition. The label on each upward triangle expresses the one-dimensional coordinate $r_1$ of the unit cell in the tilted boundary condition [Eq.~\eqref{r1_def}].}
    \label{fig:kagome_tilt}
\end{figure}

\begin{figure}[t!]
    \centering
    \includegraphics[viewport=0 0 1300 800, width=\linewidth]{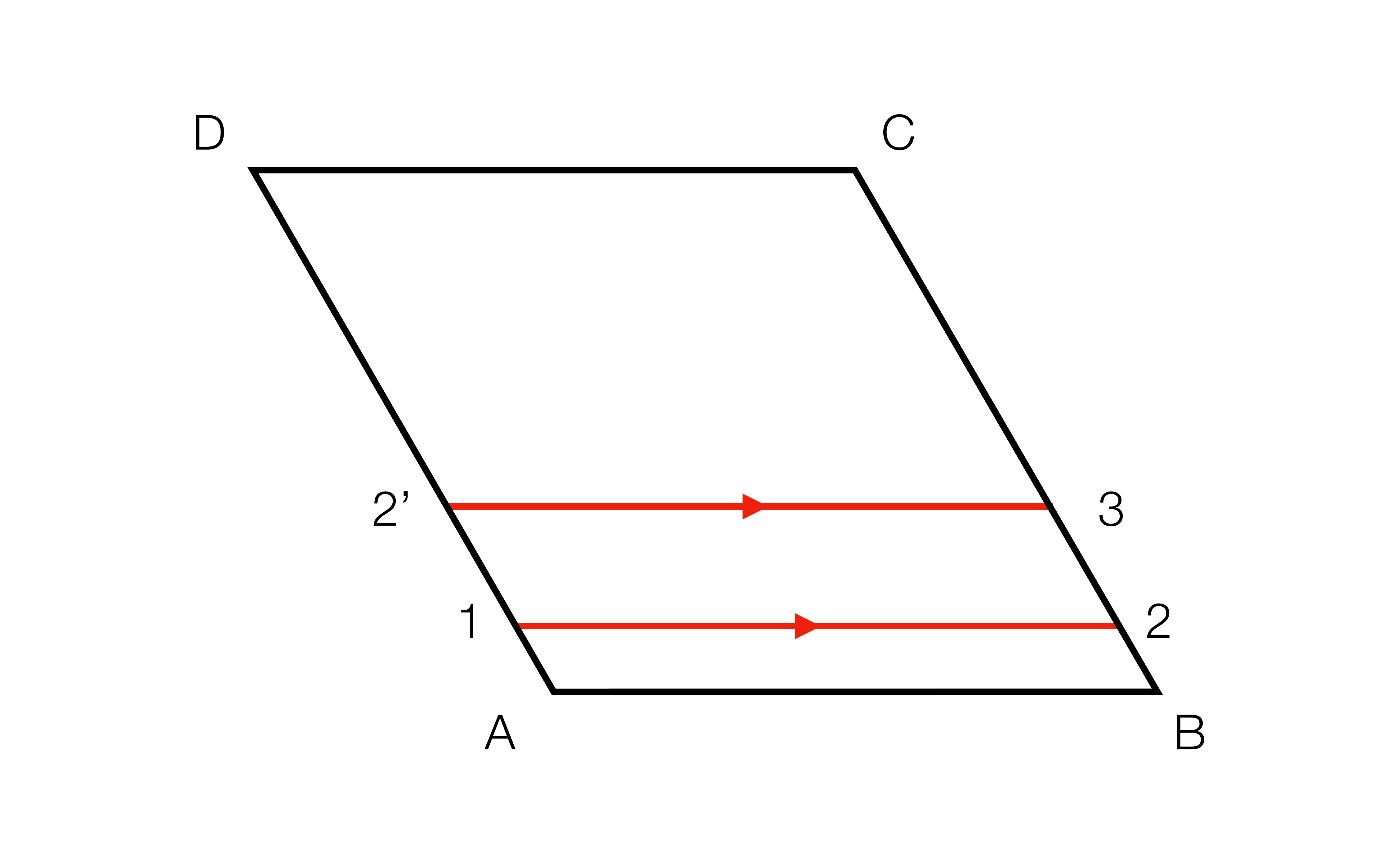}
    \caption{A schematic picture of the tilted boundary condition in a rombus finite-size cluster of the kagome lattice.
    Details of the kagome lattice are omitted, and the seam ABCD of the tilted boundary condition on which the boundary condition is imposed is shown.
    In the tilted boundary condition, a point $2$ on the right seam is identified with another point $2'$ on the left seam, where the latter is dislocated from the point $1$ by the unit vector $\bm e_2$.
    }
    \label{fig:tilt_bc}
\end{figure}

We define the tilted boundary condition on the kagome lattice.
The kagome lattice is a non-Bravais lattice whose unit cell consists of three sites in the periodic boundary condition.
Likewise, the unit cell contains three sites in the tilted boundary condition.
The one-unit translation operator $T_1$ in the $\bm e_1$ direction acts on the spin operator $\bm S_\mu(n_1,n_2)$ ($\mu=1,2,3)$ as Eq.~\eqref{T1}.
To define the tilted boundary condition, we first consider a finite-size cluster of the kagome lattice of the rhombic shape (Figs.~\ref{fig:kagome_tilt} and \ref{fig:tilt_bc}) and next take the thermodynamic limit by making the system size infinite.
The rhombic cluster breaks some of the symmetries that the infinite-size kagome lattice possesses, for example, a $C_6$ rotation symmetry whose rotation axis pierces the center of a hexagon.
The rhombic cluster is chosen because this paper is focused on an anomaly between the $\mathrm{U(1)}$ spin-rotation symmetry and the translation symmetry.
The rhombic shape excludes the effects of the $C_6$ and other symmetries on the ground-state degeneracy.

Let us define the origin $\bm R=0$ as the left bottom corner of the rhombic finite-size cluster (the center of a triangle with a label 1 in Fig.~\ref{fig:kagome_tilt}). Suppose $0\le n_1 < N_1$ and $0 \le n_2 < N_2$ for positive integers $N_1, N_2$.
On the rhombic finite-size cluster, 
the tilted boundary condition is defined as
\begin{align}
    T_1 \bm S_{\mu}\bigl(N_1-1,\, n_2\bigr) T_1^{-1} &= \bm S_{\mu}\bigl(0,\, n_2+1\bigr),
    \label{tilt}
\end{align}
for $n_2\in [0, N_2)$ and
\begin{align}
    T_1 \bm S_{\mu}\bigl(N_1-1,  N_2-1\bigr) T_1^{-1} &= \bm S_{\mu} \bigl( 0,0\bigr).
    \label{tilt_top}
\end{align}
When we reach the right seam of the system, we reenter the system from the left seam with the  dislocation by the unit vector $\bm e_2$ (Fig.~\ref{fig:tilt_bc}).
It immediately follows from Eqs.~\eqref{tilt} and \eqref{tilt_top} that the $T_2$ translation symmetry, that is, the translation symmetry in the $\bm e_2$ direction, depends on the $T_1$ one in the tilted boundary condition for a relation $T_2 = (T_1)^{N_1}$.
The aspect ratio $N_2/N_1$ of the rhombus can be arbitrary.
The total number of sites is given by
\begin{align}
    V &= 3N_1N_2
\end{align}

We can sweep all the upward triangles on the kagome lattice one dimensionally by applying $T_1$ of Eqs.~\eqref{T1}, \eqref{tilt}, and \eqref{tilt_top} repeatedly.
This path allows us to relabel the spin $\bm S_{\mu}(n_1,n_2)$ with a one-dimensional coordinate along that path as
\begin{align}
    \bm  S_{\mu}(n_1,n_2) =  \bm S_{r_1, \mu},
\end{align}
where the one-dimensional coordinate $r_1 \in [1, V/3]$ of the unit cell is related to $\bm R = n_1 \bm e_1 + n_2 \bm e_2$ through
\begin{align}
    r_1 &= 1+ n_1 + n_2N_1
    \label{r1_def}
\end{align}
The Hamiltonian \eqref{H_kagome_HAFM} in the tilted boundary condition is $T_1$-symmetric, that is, $[\mathcal H, T_1]=0$.

\subsection{Flux insertion}

\begin{table}[t!]
    \centering
    \begin{tabular}{*{7}c} \hline \hline
        \makebox[3em]{$m/S$} & \makebox[3em]{$0$} & \makebox[3em]{$1/9$} & \makebox[3em]{$1/3$} & \makebox[3em]{$5/9$} & \makebox[3em]{$7/9$} \\ \hline
        $\theta/2\pi$ & $3/2$ & $4/3$ & $1$ & $2/3$ & $1/3$ \\
        $d_{\rm m}$  & $2$ & $3$ & $1$ & $3$ & $3$ \\ \hline \hline
    \end{tabular}
    \caption{The angle $\theta$ of Eq.~\eqref{TUT} for zero magnetization and for fractional magnetizations $m/S =  (2n-1)/9$ with $n=1,2,3,4$ is listed for $S=1/2$. 
    The third row refers to the lower bound $d_{\rm m}$ of the ground-state degeneracy of the $S=1/2$ kagome antiferromagnet on those magnetization plateaus.
    }
    \label{tab:angle}
\end{table}

We insert the flux adiabatically into the Hamiltonian \eqref{H_kagome_HAFM} by replacing transverse exchange interactions,
\begin{align}
    S_{r_1,\mu}^+ S_{r'_1, \mu'}^- &+ \mathrm{H.c.} \notag \\
    &\to e^{i(r_1-r'_1) \phi/V} S_{r_1,\mu}^+ S_{r'_1,\mu'}^- + \mathrm{H.c.},
    \label{insert_flux}
\end{align}
and increase the flux amount $\phi \in \mathbb R$ slowly from zero to the unit amount, $2\pi$.
The $2\pi$ flux can be absorbed by a $\mathrm{U(1)}$ large gauge transformation,
\begin{align}
    U &= \exp\biggl(i\frac{2\pi}{V/3} \sum_{r_1=1}^{V/3} r_1 n(r_1) \biggr),
    \label{U_tilt}
\end{align}
where $n(r_1)$ is a number density of magnons at the unit cell $r_1$
\begin{align}
    n(r_1) = 3S- S_{r_1}^z,
\end{align}
with $S_{r_1}^z = \sum_{\mu=1}^3 S_{r_1,\mu}^z$.
The Hamiltonian with the flux, whatever the amount of the inserted flux is, keeps the $T_1$ symmetry and the global $\mathrm{U(1)}$ spin-rotation symmetry at the same time.
On the other hand, the translation $T_1$ and the $\mathrm{U(1)}$ large gauge transformation $U$ satisfy a relation,
\begin{align}
    T_1 U T_1^{-1} &= Ue^{i\theta},
    \label{TUT}
\end{align}
with a nontrivial angle,
\begin{align}
    \frac{\theta}{2\pi} = 3(S-m),
    \label{angle}
\end{align}
with the magnetization density per a site, $m = \sum_{r_1} S_{r_1}^z/(V/3)$.
When $\theta$ is trivial (i.e. $\theta = 0 \mod 2\pi$), the two operators $T_1$ and $U$ are commutative with each other.
Then nothing prevents the ground state from being unique and gapped.
When the system has nontrivial $\theta \not=0 \mod 2\pi$, it is forbidden to have any unique gapped ground state as follows~\cite{oshikawa_lsm}. 
Let us denote a ground state of the kagome antiferromagnet without the flux as $\ket{\Psi_0}$.
If the adiabatic flux insertion deforms smoothly $\ket{\Psi_0}$ to $\ket{\Psi'_0}$, the latter is a ground state of the kagome antiferromagnet with the unit flux.
The large-gauge-transformed $U\ket{\Psi'_0}$ is a ground state of the kagome antiferromagnet without the flux.
Note that both $\ket{\Psi_0}$ and $\ket{\Psi'_0}$ have the same eigenvalue of $T_1$ because the adiabatic flux insertion is compatible with the translation symmetry.
If $\theta$ satisfies $\theta \not=0 \mod 2\pi$, $U\ket{\Psi'_0}$ is orthogonal to $\ket{\Psi_0}$ thanks to Eqs.~\eqref{TUT} and \eqref{angle}, in other words, $U\ket{\Psi'_0}$ is a degenerate ground state or a gapless excited state of the kagome antiferromagnet without the flux.

Because the angle \eqref{angle} depends on neither $N_1$ nor $N_2$,
we can take the thermodynamic limit, $V\to +\infty$, without any ambiguity.
This well-defined thermodynamic limit is a great advantage of the tilted boundary condition over the periodic boundary condition~\cite{oshikawa_lsm, Pal_kagome}.

In the absence of the magnetic field, the model \eqref{H_kagome_HAFM} possesses the time-reversal symmetry that imposes $m=0$ unless the spontaneous ferromagnetic order is generated, which is unlikely.
Therefore, $\theta/2\pi = 3S$ follows at zero magnetic field.
When $S \in \mathbb Z+1/2$, the ground state of the spin-$S$ kagome Heisenberg antiferromagnet has either the gapless ground state or (at least) doubly degenerate gapped ground states.
The former is consistent with the $\mathrm U(1)$ Dirac spin liquid scenario~\cite{ran_u1_kagome, iqbal_u1_kagome, he_kagome_dirac} and the latter is consistent with the gapped $\mathbb Z_2$ spin liquid~\cite{Yan_z2_kagome, depenbrock_z2_kagome, jiang_z2_kagome}.
The degeneracy predicted by the relation \eqref{TUT} refers only to that by the intrinsic anomaly between the $\mathrm U(1)$ spin-rotation symmetry and the translation symmetry for a fixed filling.
The ground state can, in principle, be more degenerate than Eq.~\eqref{TUT} tells.
Therefore, the ground-state degeneracy predicted by Eq.~\eqref{TUT}, which we denote as $d_{\rm m}$,  gives the minimum possible value of the actual ground-state degeneracy.

In the presence of the magnetic field,
the spin-$1/2$ Heisenberg antiferromagnet on the kagome lattice \eqref{H_kagome_HAFM} is believed to have $1/9$, $1/3$, $5/9$, and $7/9$ magnetization plateaus.
The angles of Eq.~\eqref{angle} for those fractions of the magnetization are listed in Table~\ref{tab:angle}.
When $\theta =2\pi p/q$ with coprime integers $p$ and $q$, the following $q$ states, $\ket{\Psi_0}$, $U\ket{\Psi'_0}$, $U^2\ket{\Psi'_0}, \cdots$, and $U^{q-1}\ket{\Psi'_0}$ have the same eigenenergy in the thermodynamic limit but have different eigenvalues of $T_1$.
If the ground state is gapped, these $q$ states are $q$-fold degenerate gapped ground states.
Given this possible minimum ground-state degeneracy, of particular interest is the $1/3$ plateau where the unique gapped ground state is allowed.
In fact, Ref.~\cite{parameswaran_kagome_mott} constructed the unique gapped ground state of a model explicitly on the $1/3$ plateau without breaking any symmetry of the kagome lattice.
This is consistent with the relation \eqref{TUT}.
However, it remains obscure what allows for the unique gapped ground state on the $1/3$ plateau because the condition \eqref{TUT} with the angle \eqref{angle} only tells that the minimum number of the ground-state degeneracy allowed by the translation and the $\mathrm U(1)$ spin-rotation symmetries is $1$.
In the subsequent section, we propose one interpretation of the unique gapped ground state's appearance possible on the $1/3$ plateau from the viewpoint of an anomaly in a quasi-one-dimensional limit.

\section{Anomaly and spatial anisotropy}\label{sec:1d}

In this section, we employ the periodic boundary condition with avoiding the known problem of the size-dependent ground-state degeneracy.
The key idea is the insensitivity of a 't Hooft anomaly to spatial anisotropies.

\subsection{ insensitivity of anomalies to spatial anisotropy}

The most significant advantage of the tilted boundary condition is that it reduces the number of spins in the cross-section $M_{d-1}$ down to $O(1)$.
In the periodic boundary condition, it is $O(V^{(d-1)/d})$.
The system with $O(1)$ spins on each cross-section seems like a one-dimensional system.
In fact, we can view the kagome antiferromagnet in the tilted boundary condition as a one-dimensional quantum spin system in the periodic boundary condition where the upward triangle is one-dimensionally aligned.
The flux insertion argument is independent of whether the system is viewed as a $d$-dimensional one or a one-dimensional one.
Since the flux insertion argument picks up the anomaly between the $\mathrm{U(1)}$ symmetry and the translation symmetry, the anomaly is also independent of the viewpoint.
This observation about dimensionality motivates us to describe the anomaly of kagome quantum antiferromagnets by using one-dimensional theoretical tools.

In general, relating $d$-dimensional quantum many-body systems to one-dimensional ones is a useful idea (e.g. the coupled wire construction of topological phases~\cite{kane_cwc, lu_cwc}).
This is partly because the latter is usually much better equipped with theoretical tools than the former~\cite{giamarchi_book}.
The anomaly is no exception~\cite{furuya_wzw, cho_anomaly_lsm, yao_lsm_anomaly, tanizaki_lsm}.
In our case, however, the effective one-dimensional system inevitably contains long-range interactions that are extremely inconvenient for the anomaly argument, in particular, for the anomaly matching~\cite{thooft_matching}.
For example, a nearest-neighbor exchange interaction $\bm S_1(n_1,n_2) \cdot \bm S_2(n_1,n_2+1)$ of the Hamiltonian \eqref{H_kagome_HAFM} can be seen as an exchange interaction, $\bm S_{r_1,1} \cdot \bm S_{r_1+N_1,2}$ over the long distance $N_1$.
This is a disadvantage of the tilted boundary condition.

To relate anomalies of kagome antiferromagnets to one-dimensional physics with avoiding this inconvenience, we notice the insensitivity of anomalies to spatial anisotropies implied by the flux-insertion argument with the tilted boundary condition.
The large-gauge transformation operator \eqref{U_tilt} depends explicitly on the number of spins inside the unit cell and the lattice structure through $r_1$ but is independent of the strength of coupling constants.
In general, 't Hooft anomalies are robust against continuous variation of the Hamiltonian as long as the symmetries in question are preserved since an anomaly is identified with a surface term of a topological action of a symmetry-protected topological phase, which is manifestly topologically invariant~\cite{meng_set_surface, Jian_LSM_anomaly, thorngren_crystalline}.
We can weaken or strengthen interactions in a specific spatial direction without interfering with the flux insertion argument.
More generally, if an anomaly involves translation symmetries and on-site symmetries and is unrelated to spatial rotation symmetries, we can weaken interaction strengths in the directions perpendicular to the $\bm e_1$ direction with keeping the symmetries.
The insensitivity to the spatial anisotropy opens a way to access the accumulated knowledge about anomalies of (1+1)-dimensional quantum field theories.

This section gives particular attention to the $1/3$ plateau where the unique gapped ground state is allowed in the flux-insertion argument and is indeed constructed in a specific model~\cite{parameswaran_kagome_mott}.
We discuss that kagome antiferromagnets on the $1/3$ plateau, such as that of Ref.~\cite{parameswaran_kagome_mott}, can have the unique gapped ground state as a result of an explicit breaking of a symmetry that involves an anomaly.
This section first discusses an anomaly of a one-dimensional quantum spin system, an $S=1/2$ three-leg spin tube.
Next, using this knowledge of one-dimensional physics, 
we give an anomaly interpretation of the unique gapped ground state of the kagome antiferromagnet on the $1/3$ plateau.

\subsection{Quasi-one-dimensional limit}

In order to bridge physics in the one dimension to that on the kagome lattice,
we consider a deformed kagome antiferromagnet with the following Hamiltonian:
\begin{widetext}
\begin{align}
    \mathcal H_\delta
    &= \mathcal H_{\rm 1d} + \delta \mathcal H',
    \label{H_J1_J3} \\
    \mathcal H_{\rm 1d}
    &=J_1 \sum_{n_1,n_2} [\bm S_1(n_1,n_2) \cdot \bm S_2(n_1,n_2)
     + \bm S_2(n_1,n_2) \cdot \bm S_3(n_1,n_2) + \bm S_3(n_1,n_2) \cdot \bm S_1(n_1,n_2)]
    \notag \\
    &\quad + J_3 \sum_{n_1,n_2,\mu} \bm S_\mu (n_1,n_2) \cdot \bm S_\mu (n_1+1,n_2)
    - h \sum_{n_1,n_2,\mu} S_{\mu}^z(n_1,n_2),
    \label{H_tube} \\
    \mathcal H'
    &= J_1 \sum_{n_1,n_2} [\bm S_1(n_1,n_2) \cdot \bm S_2(n_1+1,n_2+1) + \bm S_2(n_1+1,n_2) \cdot \bm S_3(n_1,n_2) + \bm S_3(n_1,n_2+1) \cdot \bm S_1(n_1,n_2)]
    \notag \\
    &\quad +J_3 \sum_{n_1,n_2} \sum_{\mu=1,2,3} [\bm S_\mu(n_1,n_2) \cdot \{\bm S_\mu(n_1,n_2+1)  + \bm S_\mu(n_1+1,n_2+1) \}],
    \label{H'}
\end{align}
\end{widetext}
where the parameter $\delta \in [0,1]$  controls the spatial anisotropy.
The coupling constants $J_1$ and $J_3$ represent the nearest-neighbor and the third-neighbor exchange interactions when we view the model \eqref{H_J1_J3} as a two-dimensional system.

\begin{figure}[t!]
    \centering
    \includegraphics[viewport = 0 0 1400 700, width=\linewidth]{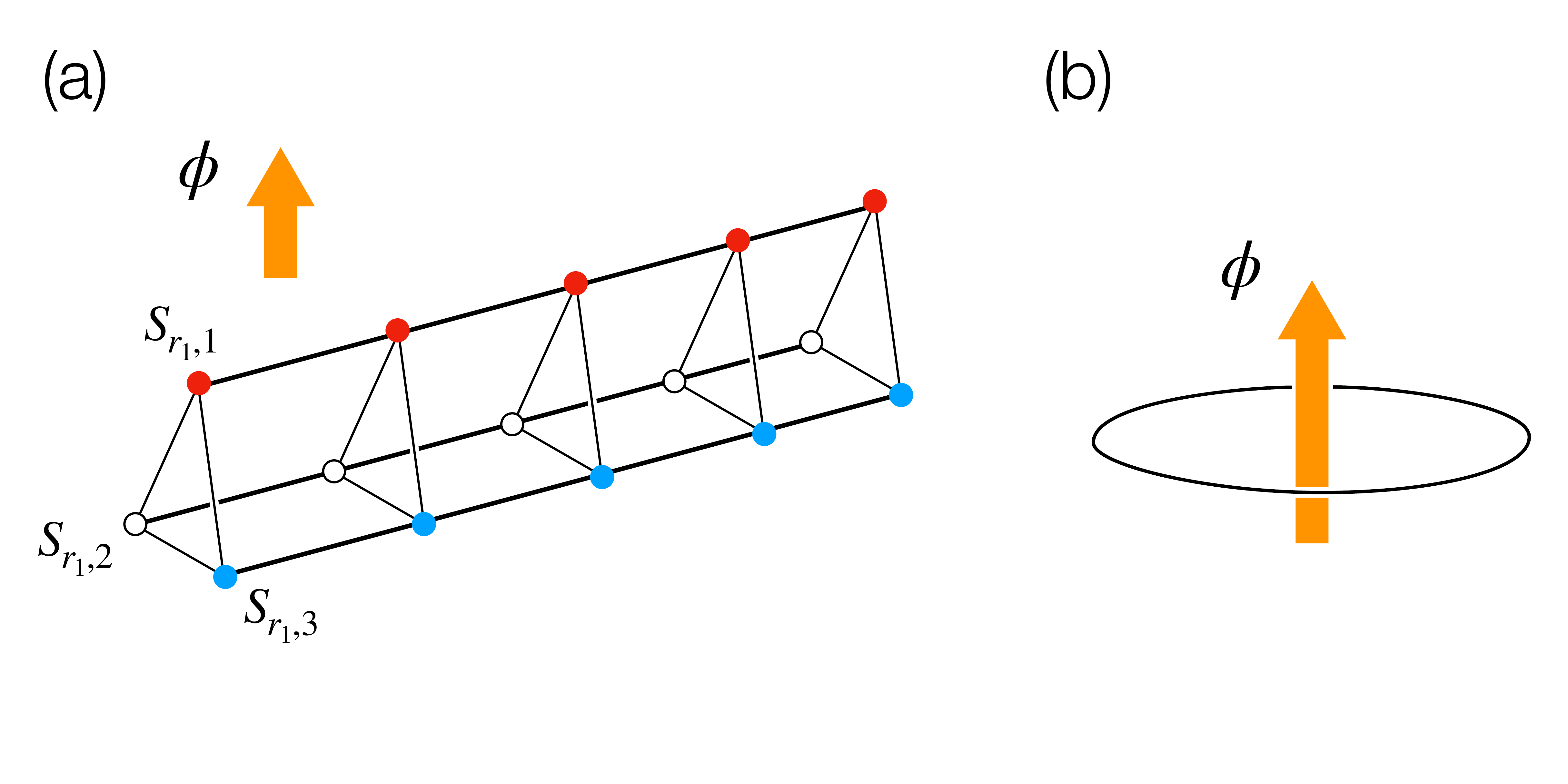}
    \caption{(a) The three-leg spin tube \eqref{H_tube} in the periodic boundary condition. 
    The periodic boundary condition is imposed on the leg direction.
    Thin and thick lines represent $J_1$ and $J_3$ interactions of Eq.~\eqref{H_tube}, respectively.
    (b) 
    The flux is inserted into the spin tube. The spin tube is regarded as a ring with a triangular cross-section.
    }
    \label{fig:tube}
\end{figure}

Let us impose the periodic boundary condition in the $\bm e_1$ and $\bm e_2$ directions.
For $\delta = 1$, all the nearest-neighbor bonds have the same strength of the exchange interaction and so do the third-neighbor ones.
For $\delta = 0$, the model \eqref{H_J1_J3} is a set of mutually independent three-leg spin tubes (Fig.~\ref{fig:tube}).
Note that the smooth change of $\delta$ keeps the $T_1$ and $T_2$ translation symmetries and the $\mathrm U(1)$ spin-rotation symmetry.
We use the periodic boundary condition to investigate the model~\eqref{H_J1_J3}.

In this section, we set $\delta =0$ for a while and discuss its 't Hooft anomaly on the $1/3$ plateau.
Later in Sec.~\ref{sec:1/3}, we will resurrect $\delta$  and discuss the anomaly in the quasi-one-dimensional limit, $0< \delta \ll 1$.

\subsection{Oshikawa-Yamanaka-Affleck condition in spin tubes}

It is now a good occasion to review the original derivation of the OYA condition in the specific case of the three-leg spin tube.
Let us consider a situation where three spin chains are weakly coupled to each other.

Each spin chain is equivalent to an interacting spinless fermion chain thanks to the Jordan-Wigner transformation~\cite{giamarchi_book}.
Let us denote an annihilation operator of the spinless fermion at a location $x$ as $\psi_{\mu}(x)$, where  $\mu=1,2,3$ are degrees of freedom that specify the leg.
In the continuum limit, $\mu$ represent internal degrees of freedom, which we call ``color'' in this paper: $\mu=1,2,3$ for red, white, and blue circles, respectively in Figs.~\ref{fig:kagome_tilt} and \ref{fig:tube}.
At low energies, the spinless fermion operator can further be split into two species:
\begin{align}
    \psi_{\mu}(x) &\approx e^{ik_Fx} \psi_{R,\mu}(x) + e^{-ik_Fx} \psi_{L,\mu}(x),
\end{align}
where $k_F>0$ is the Fermi wave number and $\psi_{R,\mu}(x)$ and $\psi_{L,\mu}(x)$ are annihilation operators of the right-moving and left-moving spinless fermions, respectively.

The magnetization process of the three-leg spin tube is described by a $\mathrm{U(1)}$ conformal field theory (CFT) of compactified bosons.
The spinless fermion operators $\psi_{R,\mu}(x)$ and $\psi_{L,\mu}(x)$ can be bosonized as 
$e^{i\phi_\mu/R} \propto \psi_{L,\mu}^\dag \psi_{R,\mu}$ and $e^{2\pi iR \theta_\mu} \propto \psi_{L,\mu}\psi_{R,\mu}$.
Here, $R>0$ is the compactification radius of those bosons.
The $\mathrm{U(1)}$ CFT is written in two compactified $\mathrm{U(1)}$ bosons $\phi = \sum_{\mu=1}^3\phi_\mu$ and $\theta = \sum_{\mu=1}^3\theta_\mu$.

The Fermi wave number $k_F$ is related to the magnetization per site $m$ as
\begin{align}
    k_F = \frac{\pi}{a} \biggl( \frac 12 - m \biggr),
\end{align}
where $a$ is the lattice spacing of the spin chain.
The $T_1$ translation $x \to x+a$ along the leg effectively turns into an on-site symmetry, $\psi_{R,\mu}(x) \to e^{ik_Fa} \psi_{R,\mu}(x)$ and $\psi_{L,\mu}(x) \to e^{-ik_Fa} \psi_{L,\mu}(x)$ in the continuum limit $a \to 0$.
In the boson language, the $T_1$ translation $x \to x+a$ affects only $\phi_\mu$:
\begin{align}
    \phi_\mu  &\to \phi_\mu + 2ak_FR, \\
    \phi &\to \phi + 6ak_FR.
\end{align}
When $6ak_F \in 2\pi \mathbb Z$, a perturbation $\cos(\phi/R)$ of the $\mathrm{U(1)}$ CFT is permitted by the translation symmetry.
Note that cosines and sines of $\theta$ are forbidden by the $\mathrm{U(1)}$ spin-rotation symmetry.
If $\cos(\phi/R)$ is relevant in the sense of the renormalization group, this cosine interaction opens the gap without any spontaneous symmetry breaking.
If it is irrelevant, the ground state is gapless because it is the most relevant operator allowed by the symmetries.
The condition $6ak_F \in 2\pi \mathbb Z$ is rephrased as
\begin{align}
    3(S-m)\in \mathbb Z,
    \label{OYA}
\end{align}
with $S=1/2$.
The relation \eqref{OYA} is the OYA condition in the case of the three-leg spin tube.
It is straightforward to derive the same condition \eqref{OYA} for higher spin quantum numbers $S> 1/2$.

When $6ak_F \in 2\pi(\mathbb Z+ p/q)$ with coprime positive integers $p$ and $q$, the most relevant symmetric interaction in the $\mathrm{U(1)}$ CFT is $\cos(q\phi/R)$.
If the ground state is gapped, the ground state is at least $q$-fold degenerate by breaking the translation symmetry $\phi \to \phi + 2\pi p R/q \mod 2\pi R$ spontaneously.
We thus reached the same conclusion as that derived from Eqs.~\eqref{TUT} and \eqref{angle}.
However, this is not the end of the story.

\subsection{Color constraints on the $1/3$ plateau}\label{sec:color}

$S=1/2$ three-leg spin tubes possess the $1/3$ plateau~\cite{okunishi_tube_chiral, plat_tube, yonaga_tube_twisted}.
According to the OYA condition \eqref{OYA}, the $S=1/2$ three-leg spin tube can, in principle, have the unique gapped ground state on the $1/3$ magnetization plateau since $3(S-m) = 1\in \mathbb Z$.
On the other hand, to the best of our knowledge, the unique gapped ground state has not yet been reported for the simple three-leg spin tube with the Hamiltonian equivalent to Eq.~\eqref{H_tube} on the $1/3$ magnetization plateau~\cite{okunishi_tube_chiral, yonaga_tube_twisted}.
This is not a coincidence.
The unique gapped ground state is indeed forbidden, as we show below.

We reuse the spinless fermion picture.
When three spinless fermion chains are gapless interacting Dirac ones decoupled from each other, the system has the $\mathrm{U(3)}$ symmetry.
Interchain interactions in the spin tube reduce this symmetry.
The charge of the spinless fermion corresponds to the magnetization of the spin tube.
On the $1/3$ magnetization plateau where the total $S^z$ is frozen, magnetic excitations can be gapped without breaking any symmetry, as the OYA condition \eqref{OYA} implies.
The remaining degrees of freedom, which are the color,  are effectively described by a perturbed $\mathrm{SU(3)}$ WZW theory~\cite{affleck_bosonization}.
At the fixed point, the $\mathrm{SU(3)}$ WZW theory has a global $\frac{\mathrm{SU(3)}_R\times \mathrm{SU(3)}_L}{\mathbb Z_3}$ symmetry~\cite{ohmori_flag}, where $V_{R/L} \in \mathrm{SU(3)}_{R/L}$ act on the field $g \in \mathrm{SU(3)}$ of the WZW theory as $g \to V_L gV_R^\dag$.
The symbols $R$ and $L$ represent the right-moving and the left-moving parts of particles of the WZW theory.
In terms of spinless fermions $\psi_{R/L, \mu}$, a $(\mu,\nu)$ component, $g_{\mu\nu}$, of the $\mathrm{SU(3)}$ matrix $g$ can be represented as $g_{\mu \nu} \propto \psi_{L,\mu}^\dag \psi_{R,\nu}$~\cite{Affleck_sun}.
Generally, the $R$ and $L$ degrees of freedom are coupled to each other by intrachain and interchain interactions when the field theory deviates from the fixed point.
$\mathrm{SU(3)}_R\times \mathrm{SU(3)}_L$ is reduced to a single $\mathrm{SU(3)}$ with $V_R=V_L$, away from the fixed point.
Then, the global symmetry is reduced to $\mathrm{PSU(3)}$.

We did not include this $\mathrm{SU(3)}$ WZW theory in the previous subsection.
The $\mathrm{SU(3)}$ WZW theory is actually crucial on the $1/3$ magnetization plateau because of the following reason.
The Fermi wave number $k_F$ is given by $k_F = \pi/3a$ on the $1/3$ plateau.
The field $g_{\mu\nu} \propto \psi_{L,\mu}^\dag \psi_{R,\nu}$ transforms under the $T_1$ translation as
\begin{align}
    g \to e^{2\pi i/3} g.
    \label{z3_trn}
\end{align}
The $T_1$ translation symmetry is turned into the on-site $\mathbb Z_3$ symmetry \eqref{z3_trn}.
The $\mathrm{SU(3)}$ WZW theory has an anomaly between the $\mathbb Z_3$ symmetry \eqref{z3_trn} and the $\mathrm{PSU(3)}$ symmetry~\cite{yao_lsm_anomaly}.
According to Ref.~\cite{yao_lsm_anomaly}, the $\mathbb Z_3$ symmetry \eqref{z3_trn} and the $\mathrm{PSU(3)}$ symmetry give an LSM index $\mathcal I_3$ that equals the number of the Young-tableau box per unit cell.
The LSM index is a quantity related to the ground-state degeneracy.
If and only if $\mathcal I_3 =0 \mod 3$, the ground state can be unique and gapped.
Otherwise, the ground state is either gapless or gapped with at least $q$-fold degeneracy.
Here, $q$ is given by 
\begin{align}
    q= \frac{3}{\operatorname{gcd}(\mathcal I_3, 3)},
    \label{gsd}
\end{align}
where $\operatorname{gcd}(n_1,n_2)$ is the greatest common divisor of two integers $n_1$ and $n_2$.
The $3\times 3$ matrix $g$ belongs to either the fundamental or the conjugate representation of $\mathrm{SU(3)}$.
The number of the Young tableau of the fundamental (conjugate) representations contains one box (two boxes, respectively).
Hence, the three-leg spin tube on the $1/3$ plateau has the LSM index $\mathcal I_3=1$ or $\mathcal I_3=2$, both of which lead to $q=3$.
The spin tube's ground state on the $1/3$ plateau is either gapless or gapped with at least three-fold degeneracy.
The gapless ground state is indeed possible on the $1/3$ magnetization plateau, which is a liquid state of chirality degrees of freedom~\cite{okunishi_tube_chiral, plat_tube}.

We saw that the translation and the $\mathrm{PSU(3)}$ color symmetries forbid the unique gapped ground state on the $1/3$ plateau.
However, the $\mathrm{PSU(3)}$ color symmetry is too large to be naturally realized in three-leg spin tubes.
In fact, in general, spin tubes have a $\mathbb Z_3$ color-rotation symmetry instead of $\mathrm{PSU(3)}$.
It is desirable to reduce the symmetries that forbid the unique gapped ground state as much as possible.

We can reduce the color symmetry relevant to the ground-state degeneracy by referring to another quantum field theory, a (1+1)-dimensional $\mathrm{SU(3)/U(1)^2}$ nonlinear sigma model~\cite{lajko_su3, tanizaki_lsm}.
This nonlinear sigma model, as well as the $\mathrm{SU(3)}$ WZW theory, is also an effective field theory of $\mathrm{SU(3)}$ spin chains~\cite{lajko_su3, tanizaki_lsm, yao_lsm_anomaly}.
Though the $\mathrm{SU(3)/U(1)^2}$ model is not directly related to our spin tube~\eqref{H_tube},
we consider this nonlinear sigma model for its relation to the $\mathrm{SU(3)}$ WZW theory;
the low-energy limit of the (1+1)-dimensional $\mathrm{SU(3)/U(1)^2}$ nonlinear sigma model perturbed by local interactions is described by the $\mathrm{SU(3)}$ WZW theory~\cite{ohmori_flag}.
According to the anomaly matching argument~\cite{thooft_matching}, the $\mathrm{SU(3)/U(1)^2}$ nonlinear sigma model and the $\mathrm{SU(3)}$ WZW theory share the mixed 't Hooft anomaly in common.
In fact, Ref.~\cite{tanizaki_lsm} shows that the $\mathrm{SU(3)/U(1)^2}$ nonlinear sigma model has an anomaly between the $\mathrm{PSU(3)}$ symmetry and the one-unit translation symmetry.
This is consistent with the result of Ref.~\cite{yao_lsm_anomaly}.
Reference~\cite{tanizaki_lsm} further argued that the nonlinear sigma model has an anomaly between
a finite subgroup $\mathbb Z_3\times \mathbb Z_3$ of $\mathrm{PSU(3)}$ and the translation symmetry.
Note that $\mathbb Z_3\times \mathbb Z_3$ acts on the sigma fields projectively~\cite{tanizaki_lsm}.

The $\mathbb Z_3\times \mathbb Z_3$ group is related to the following matrices,
\begin{align}
    M_1 &=
    \begin{pmatrix}
    1 & 0 & 0 \\
    0 & \omega & 0 \\
    0 & 0 & \omega^2
    \end{pmatrix}, \qquad
    M_2 = 
    \begin{pmatrix}
        0 & 1 & 0 \\
        0 & 0 & 1 \\
        1 & 0 & 0
    \end{pmatrix}.
\end{align}
An isomorhpism $H:\mathrm{SU(3)} \to \mathrm{PSU(3)}$ maps $M_n$ for $n=1,2$  to $H(M_n)$ that generate the $\mathbb Z_3\times \mathbb Z_3$.
Note that $M_1$ is a gauge symmetry of the $\mathrm{SU(3)/U(1)^2}$ nonlinear sigma model, which does not forbid any interaction that perturbs the nonlinear sigma model.
By contrast, $M_2$ forbids some interactions.
It acts on the $g$ field as $M_2 g M_2^\dag$.
In other words, $M_2$ represents the $\mathbb Z_3$ color-rotation symmetry, which is essential in what follows.

The $\mathbb Z_3$ color-rotation symmetry forbids interactions that treat the legs of the spin tube unequally.
By imposing the $\mathbb Z_3$ color-rotation symmetry, the $\mathrm{U(1)}$ spin-rotation symmetry, and the translation symmetry on the spin tube, we can forbid the system \eqref{H_tube} from having the unique gapped ground state on the $1/3$ plateau.
On the other hand, this anomaly is absent in generic kagome antiferromagnets such as Eq.~\eqref{H_kagome_HAFM}, as we show below.

\subsection{Unique gapped ground state on the $1/3$ plateau}\label{sec:1/3}

While the anomaly argument in Sec.~\ref{sec:color} is already nontrivial as quantum physics in one dimension, it also gives interesting feedback to the original two-dimensional problem.
Let us look back on the kagome antiferromagnet \eqref{H_tube}.
As we mentioned, the anomaly in the two-dimensional system is inherited by the quasi-one-dimensional system where, in our case, three-leg spin tubes are weakly coupled with short-range interactions.
Recall that the periodic boundary conditions are imposed on the $\bm e_1$ and the $\bm e_2$ directions.
Hence, the anomaly argument in the quasi-one-dimensional limit reversely gives us a glimpse of the anomaly in the isotropic two-dimensional case.
Let us suppose that these intertube interactions respect the $\mathbb Z_3$ color-rotation symmetry, the $\mathrm U(1)$ spin-rotation symmetry, and the $T_1$ and $T_2$ translation symmetries.

The unique gapped ground state is forbidden under those symmetries of intertube interactions.
When the single spin tube has degenerate gapped ground states with a spontaneous symmetry breaking, the weak enough symmetric intertube interactions keep the gap open and do not lift the degeneracy.

On the other hand, when the single spin tube has the gapless ground state on the $1/3$ plateau, that is, the chirality liquid state~\cite{okunishi_tube_chiral}, a weak symmetric intertube interaction can induce a quantum phase transition.
This phase transition occurs even if the coupling constant of the intertube interaction is infinitesimal~\cite{schulz_interchain, furuya_interchain}.
This is because the chirality liquid state is a Tomonaga-Luttinger liquid.
At zero temperature, the Tomonaga-Luttinger liquid state has a divergent susceptibility of the order parameter~\cite{giamarchi_book}.
This quantum phase transition drives the chirality liquid state into an ordered phase where the $\mathbb Z_3$ symmetry is spontaneously broken.
After all, the unique gapped ground state is forbidden in both cases.

Now we go back to the $J_1-J_3$ model \eqref{H_J1_J3} of the kagome antiferromagnet.
We showed that the model with $\delta=0$ has the anomaly between the $\mathbb Z_3$ color-rotation symmetry and the $T_1$ translation symmetry on the $1/3$ plateau.
As soon as $\delta$ is turned on,
the $\mathbb Z_3$ symmetry breaks down because of the first three terms of Eq.~\eqref{H'} with the coupling $J_1$ break the $\mathbb Z_3$ color-rotation symmetry.
On the basis of the anomaly argument, we can expect that the absence of the $\mathbb Z_3$ color-rotation symmetry makes the unique gapped ground state possible on the $1/3$ plateau of the kagome antiferromagnet.

In the end, we caught a glimpse of the anomaly of the kagome antiferromagnet on the $1/3$ plateau with the high $\mathbb Z_3$ color symmetry by fine-tuned parameters.
However, we also found no impediment to the possibility of the unique gapped ground state in general kagome geometry because the $\mathbb Z_3$ color-rotation symmetry is broken unless parameters of the Hamiltonian are fine-tuned.

Our argument shows that there is no unique gapped ground state on the $1/3$ plateau of a model with the $\mathbb Z_3$ color-rotation symmetry, the $\mathrm U(1)$ spin-rotation symmetry, and the $T_1$ and $T_2$ translation symmetries beyond the quasi-one-dimensional limit.
Unfortunately, it remains challenging to confirm the existence of this anomaly directly in the isotropic two-dimensional case.
We leave it for future works.

\subsection{Reduction of the unit cell}
\label{sec:reduc}

To close this section, we need to mention an unlikely possibility of reducing the number of spins in the unit cell.
In this paper, we took an upward triangle as a unit cell as a natural choice.
We can choose, in principle, a unit cell with any other shape.
One might expect that regardless of the unit cell's shape, the unit cell of the kagome lattice will have three spins inside the unit cell.

This statement is actually nontrivial when we take into account general closed boundary conditions.
For example, antiferromagnets on the checkerboard lattice have two spins inside the unit cell under the periodic boundary condition or the tilted boundary condition but have only one spin under the spatially twisted boundary condition~\cite{furuya_checkerboard}.
This reduction of the number of spins in the unit cell was a key to show the impossibility of the unique gapped ground state in the checkerboard antiferromagnet at zero magnetic field in Ref.~\cite{furuya_checkerboard}. 
If we could construct a symmetric closed boundary condition with a unit cell containing only one spin on the kagome lattice, the lower bound $d_{\rm m}$ shown in Table.~\ref{tab:angle} would be raised.

The bottom line is that such closed boundary conditions with the reduced unit cell can indeed be constructed, but they break the $T_1$ translation symmetry.
These closed boundary conditions with the reduced unit cell do not qualify as the symmetric closed boundary condition in the kagome lattice case different from the checkerboard lattice.
We construct those closed boundary conditions in Appendix~\ref{app:C3}.
We thus close this section by concluding that $d_{\rm m}$ in Table.~\ref{tab:angle} gives the maximum value of the lower bound of the ground-state degeneracy under the $\mathrm{U(1)}$ spin-rotation symmetry and the translation symmetry.

\section{Summary}\label{sec:summary}

We discussed magnetization plateaus of geometrically frustrated quantum antiferromagnets on the kagome lattice from the viewpoint of the OYA condition in one symmetric boundary condition, the tilted boundary condition~\cite{yao_lsm_boundary}.
Here, the OYA condition was derived in a form independent of the aspect ratio $N_2/N_1$ of the rhombic finite-size cluster of the kagome lattice.
Using the flux insertion argument with the tilted boundary condition,
we showed that the ground states on the $1/9$, $5/9$, and $7/9$ plateaus of spin-$1/2$ kagome antiferromagnets are at least threefold degenerate.
The $1/3$ plateau, different from the other plateaus, can host a unique gapped ground state, as was explicitly demonstrated before in a specific model~\cite{parameswaran_kagome_mott}.

To foster a better understanding of the $1/3$ plateau, we gave our attention to the insensitivity of the anomaly to spatial anisotropies when no spatial rotation symmetry is required for the appearance of the anomaly.
Investigating the anomaly in the quasi-one-dimensional limit is thus expected to help understand the anomaly of higher-dimensional systems.
We exemplified the usefulness of this idea in the kagome antiferromagnet on the $1/3$ magnetization plateau.

The $1/3$ magnetization plateau has one special significance in this viewpoint.
The 't Hooft anomaly argument of (1+1)-dimensional quantum field theories clarified one characteristic feature of the $1/3$ plateau of the kagome lattice:  
The simple three-leg spin tube \eqref{H_tube} with the $\mathbb Z_3$ color symmetry cannot have the unique gapped ground state on the $1/3$ magnetization plateau because of the anomaly between this symmetry and the translation symmetry.
This impossibility of the unique gapped ground state is not predicted in the original OYA condition~\cite{oya_plateau}.
This is because the OYA condition refers to the spin degrees of freedom, but our argument here involves the color degrees of freedom.
On the other hand, when we adopt this argument to the kagome antiferromagnet, we find that this system can have the unique gapped ground state because the kagome lattice breaks the $\mathbb Z_3$ color rotation symmetry unless the Hamiltonian is fine-tuned.
We thus conclude that the absence of the $\mathbb Z_3$ color symmetry enables the unique gapped ground state on the $1/3$ plateau of kagome antiferromagnets in general.

\section*{Acknowledgments}

The authors are grateful to Yohei Fuji, Akira Furusaki, and Takuya Furusawa for stimulating discussions.
This work was supported by KAKENHI Grant Nos. JP16K05425 and JP20K03778 from the Japan Society for the Promotion of Science.
S.C.F is supported by a Grant-in-Aid for Scientific Research on Innovative Areas “Quantum Liquid Crystals” (Grant No. JP19H05825).

\appendix

\section{Closed boundary condition with spatial rotations}
\label{app:C3}

\subsection{Reduction of the number of spins}

We saw that the kagome lattice does not have the $\mathbb Z_3$ color symmetry in the tilted boundary condition without fine-tuning parameters.
On the other hand, the absence of the $\mathbb Z_3$ color-rotation symmetry sounds strange because the three colors of the kagome lattice look equivalent in the isotropic thermodynamic limit.
In this appendix, we construct a boundary condition to respect this $\mathbb Z_3$ color-rotation symmetry of the kagome lattice.
However, the translation symmetry is broken with this boundary condition, as we show later.

We have another motivation to consider such a strange boundary condition, as mentioned in Sec.~\ref{sec:reduc}.
We will see that the boundary conditions constructed in what follows contain only one spin inside the unit cell.

\begin{figure}
    \centering
    \includegraphics[viewport=0 0 800 800, width=0.5\linewidth]{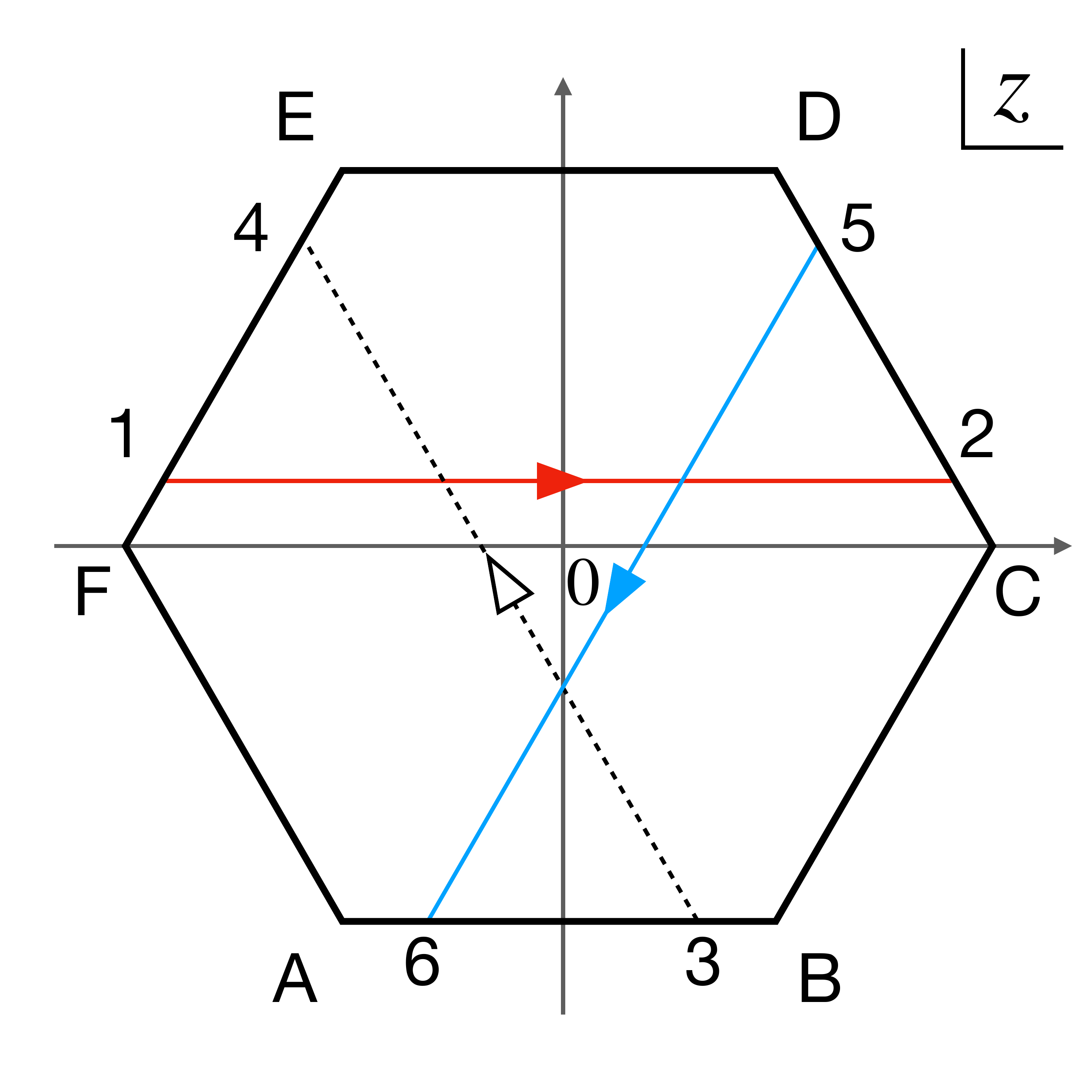}
    \caption{The $C_3$-rotated boundary condition. The kagome lattice is put on the complex plane $\mathbb C$ whose origin is $C_3$-invariant. Translation rules \eqref{T1_c3_bulk} in the bulk and \eqref{T1_c3_seam} on the seam enables us to follow a closed path $1\to 2 \to 3 \to 4 \to 5 \to 6 \to 1$, starting from a point $1$ on the left seam AFE.}
    \label{fig:c3_bc}
\end{figure}

We can construct two closed boundary conditions with which the unit cell contains only one spin.
One of these boundary conditions is referred to as a $C_3$-rotated boundary condition in this paper because it goes with the $C_3$ spatial rotation on the system's seams (Fig.~\ref{fig:c3_bc}).
Another is referred to as a tilted $C_3$-rotated boundary condition, a combination of the $C_3$-rotated boundary condition and the tilted boundary condition.

\subsection{Rotated boundary conditions}

The $C_3$-rotated boundary condition is defined as follows.
Let us consider a hexagonal finite-size cluster of the kagome lattice and denote the location of each site as a single complex variable $z$.
We define the origin $z=0$ as the $C_3$-rotation invariant point.
Accordingly, we relabel the spin operator as
$\bm S_\mu(z)$ for $\mu=1,2,3$.
The translation $T_1$ in this boundary condition is defined as
\begin{align}
    T_1 \bm S_\mu(z)T_1^{-1} &= \bm S_\mu(z+2\omega^{\mu-1}),
    \label{T1_c3_bulk}
\end{align}
with $\omega=\exp(2\pi i/3)$ if a segment that connects $z$ and $z+ 2\omega^{\mu-1}$ does not overpass the seam ABCDEF of Fig.~\ref{fig:c3_bc}.
If it overpasses the seam, we perform a $C_3$ rotation,
\begin{align}
    T_1 \bm S_\mu(z_{\rm seam}) T_1^{-1} &=  \bm S_{\mu+1}(-\omega^{2-\mu} \bar z_{\rm seam}),
    \label{T1_c3_seam}
\end{align}
where $z_{\rm seam}$ is the complex coordinate on the seam of the system and $\bm S_4(z) = \bm S_1(z)$.

\begin{figure}[t!]
    \centering
    \includegraphics[viewport = 0 0 900 750, width=\linewidth]{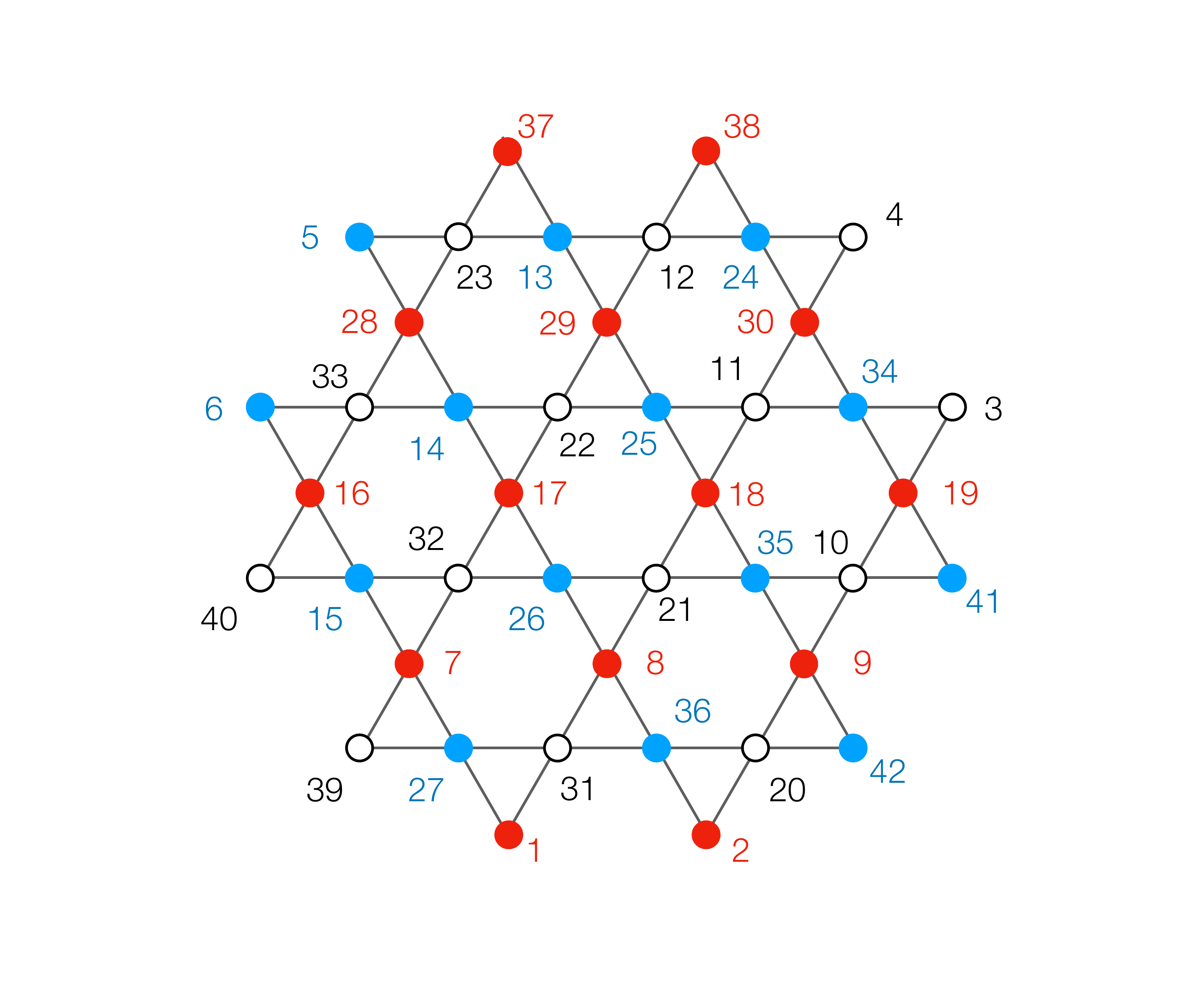}
    \caption{The $42$-site kagome lattice in the tilted $C_3$-rotated boundary condition. Every site is labeled by a single number that corresponds to the one-dimensional coordinate $r'_1$. The $T_1$ translation increases $r'_1$ by one.}
    \label{fig:kagome_rot}
\end{figure}

This boundary condition is depicted in Fig.~\ref{fig:c3_bc}.
Let us apply the translation repeatedly and return to the starting point, say, the point $1$.
We denote the complex coordinate of the point $n$ of Fig.~\ref{fig:c3_bc} as $z_n$.
These complex coordinates are related through
\begin{align}
    (z_1, z_3, z_5) &= \omega (z_5, z_1, z_3), \\
    (z_2, z_4, z_6) &= \omega (z_6, z_2, z_4), \\
    (z_1, z_3, z_5) &= (-\omega^2 \bar z_6, -\omega \bar z_2, -\bar z_4),
\end{align}
where $\bar z$ is the complex conjugate of $z$.
The point $1$ is located at the left seam of the system.
If we denote the real part of $z_1$ as $-\ell$, spin operators $\bm S_\mu(z_n)$ are related to each other in accordance with Eqs.~\eqref{T1_c3_bulk} and \eqref{T1_c3_seam}:
\begin{align}
    (T_1)^{\ell} \bm S_1(z_1) (T_1)^{-\ell} &= \bm S_1(z_2), \\
    T_1 \bm S_1(z_2) T_1^{-1} &= \bm S_2(z_3), \\
    (T_1)^{\ell} \bm S_2(z_3) (T_1)^{-\ell} &= \bm S_2(z_4), \\
    T_1 \bm S_2(z_4) T_1^{-1} &= \bm S_3(z_5), \\
    (T_1)^{\ell} \bm S_3(z_5) (T_1)^{-\ell} &= \bm S_3(z_6), \\
    T_1 \bm S_3(z_6) T_1^{-1} &= \bm S_1(z_1).
\end{align}
Note that $\ell$ is a positive integer that depends on $z_1$.
Repeated operations of $T_1$, thus bring us back to the starting point, $z_1$, eventually.

To insert the flux, we introduce a tilt to the $C_3$-rotated boundary condition.
Namely, we modify Eq.~\eqref{T1_c3_seam} to
\begin{align}
    T_1 \bm S_\mu (z_{\rm seam}) T_1 ^{-1} &= \bm S_{\mu+1} (-\omega^{2-\mu}\bar z_{\rm seam} + i\sqrt 3 \delta_{\mu,3}),
    \label{T1_tc3_seam}
\end{align}
where $\delta_{a,b}$ is the Kronecker delta.
We call the boundary condition \eqref{T1_tc3_seam} as a tilted $C_3$-rotated boundary condition.

Operations \eqref{T1_c3_bulk} and \eqref{T1_tc3_seam} define a path $\mathcal C$ along which we can sweep every spin on the kagome lattice once and only once.
Let us define the starting point as a site at the left bottom corner (Fig.~\ref{fig:kagome_rot}).
We can claim that adiabatically inserted unit flux is eliminated by a large gauge transformation $U_{\rm rot}$ defined as
\begin{align}
    U_{\rm rot}
    &= \exp\biggl( i\frac{2\pi}{V} \sum_{r'_1=1}^V r'_1 n(r'_1) \biggr),
    \label{U_rot}
\end{align}
where $r'_1$ is the one-dimensional coordinate corresponding to $z$ along the path $\mathcal C$.
However, the Hamiltonian on the kagome lattice in this boundary condition inevitably breaks the translation symmetry along the path $\mathcal C$.
In other words, the $T_1$ operator defined as Eqs.~\eqref{T1_c3_bulk} and \eqref{T1_tc3_seam} leads to $[\mathcal H, T_1]\not=0$.
The absence of the translation symmetry is evident from, for instance, the inequivalence of sites $1$ and site $7$ of Fig.~\ref{fig:kagome_rot}.
The spin on the latter site has nearest-neighbor exchange interaction between four neighboring sites, but the spin on the former site has those between only two neighboring sites.
Of course, we did not exclude rigorously at all the possibility of the symmetric closed boundary condition with which the unit cell contains one spin.
However, it is also certain that almost no room is left for this possibility.
We thus conclude that it is highly unlikely that the number of spins in the unit cell is reduced without any explicit symmetry breaking.

\bibliography{ref.bib}

\end{document}